\documentclass[aps,prl,onecolumn,showpacs,preprintnumbers,amsmath,amssymb]{revtex4-1}
\usepackage[colorlinks=true,citecolor=red,filecolor=green,linkcolor=blue,pdfnewwindow=true]{hyperref}
\usepackage{amsmath} 
\usepackage{amssymb}
\usepackage[version=4]{mhchem}
\usepackage{color}
\usepackage{graphicx}
\usepackage{makeidx}
\usepackage{bm}
\usepackage[title]{appendix} 
\usepackage{geometry} 
\usepackage{float}
\usepackage{caption}
\usepackage{hyperref}
\usepackage[normalem]{ulem}
\usepackage{natbib}
\usepackage[dvipsnames]{xcolor}
\usepackage[english]{babel}
\usepackage[autostyle]{csquotes}
\usepackage{epigraph}
\usepackage{threeparttable}
\usepackage{multirow}
\usepackage{booktabs}
\usepackage{subcaption}

\hypersetup{urlcolor=blue}

\makeatletter
 
\newcommand{\Rmnum}[1]{\expandafter\@slowromancap\romannumeral #1@}
\makeatother

\begin{document}

\title {On study of cell proliferation and diffusion using nonlinear transforms of heat equation solutions}

\author{Preet Mishra}
\author{Shyam Kumar}
\author{Sapna Ratan Shah}
\author{R.~K.~Brojen Singh}
\email{brojen@jnu.ac.in}
\affiliation{School of Computational $\&$ Integrative Sciences, Jawaharlal Nehru University, New Delhi-110067, India.}

\begin{abstract}
\noindent Cell proliferation and diffusion can be modeled through reaction-diffusion systems describing the space-time evolution of a density variable. In this work, we present non-linear transformations of heat equation solutions to model cellular growth and diffusion using a Richards growth function. The solutions are obtained by using two-variable Hermite Polynomials. We estimated the parameters of the Richards function by comparing the model solutions with the experimentally observed data from scratch assays. To check robustness of the parameters estimated, we used a fractional Brownian Motion (fBM) field type noise with given Hurst exponent (H) for minimizing residual errors. We found that the parameters can be robustly estimated and match with previous estimations. Further, we also confirmed that the spatial-temporal patterns of cell density in the experiments can be well described by the solutions developed using the proposed nonlinear transforms method.  \\

{\noindent}{{\it \textbf{Keywords:}}  \textbf{Hopf-Cole transforms}; \textbf{Richards model ODE}; \textbf{Parameter estimation}; \textbf{fBM noise}; \textbf{Hermite Polynomials} }

\end{abstract}

\maketitle


\section{Introduction}

{\noindent}Cells are extremely complex living systems. Since their growth involves the coordination of large number of diverse proteins and their physiological properties driven by inter and intracellular processes, it is quite difficult to study and model it. However, there have been many simple to complex mathematical models to study it \cite{Andrusyak}, but still we need to go a long way to address various issues so that the proposed models can correlate with the actual experimental observations. In general, the modeling of growth phenomena often leads to the representation of the system by a variable $u(t) \in [0,1]$ representing normalized (with respect to a carrying capacity or saturation level)  cell population. A simple prototypical growth model is of the form,
\begin{equation}
    \frac{du(t)}{dt}=F[u(t)]
    \label{grow}
\end{equation}
Since, this dynamics of the growth is associated with complex diffusion processes of the cells, this equation \eqref{grow} can now be represented by a complex nonlinear prototypical, dimensionlss reaction-diffusion \cite{kpp,fisher} equation (RDE) of the form,
\begin{eqnarray}
\Omega( u_t,u_{\textbf{x}\textbf{x}},u_\textbf{x},F(u),\textbf{x},t) =0
\label{rde}
\end{eqnarray}
where, $u : \mathbb{R}^d \times [0,\infty) \rightarrow [0,1]$ is the normalized density, $\textbf{x} \in \mathbb{R}^d$, $t \in [0,\infty) $ where $F(u(\textbf{x},t))$ denotes the function for all interactive mechanisms of the system under consideration. For detailed discussion and rigorous derivation of these equations from microscopic interactions, we refer to the works in \cite{kardar,derrida,panja}.\\

{\noindent}Now, we focus on a special class of grow models of cell proliferation and diffusion. The rationale for this class of models for cellular processes involving growth and cellular diffusion can be traced back to the classical works in \cite{canosa1,canosa2}, where, the Fisher type nonlinearity ($F(u) = u(1-u)$) was studied in details. Since then numerous works have explored this theme from which we give a small sample. The works  in \cite{mccue,cai,scribner} describe the cross-talk between experimental data analysis and modeling. Further the works in \cite{rutter1,rutter2} give the role of density dependent diffusion in various cellular processes. For a detailed reviews and an extensive bibliography on these themes we refer to \cite{baker3}.\\

\noindent These kind of growth models derived from stochastic processes were also studied in \cite{kardar} through the moments of the density variable. There, the authors assumed that growth and diffusion act in alternating steps, leading to the emergence of scaling exponents between these moments. Independently, a theoretical approach in \cite{baker2} introduced a multi-stage model to explain cell growth and diffusion, distinguishing between phases of the cell cycle. A key assumption in their framework was that diffusion occurs only during specific phases. Through a traveling wave ansatz, \cite{baker2} revealed a connection between PDEs with density-dependent diffusion and multi-stage sub-population models i.e. coupled PDEs. The fact that these two independent modeling perspectives—one emphasizing moment scaling due to alternating dynamics \cite{kardar}, and the other highlighting density-dependent diffusion and proliferation \cite{baker2} in a structured cell cycle falls in the same class of solutions as the ones derived below form the core motivation for this work.\\

\noindent In contrast to both approaches, we directly examine the role of non-linear transformations to construct solutions where growth and diffusion act as operators. The technique we used in this work is generally known as nonlinear invertible transformations of heat equation solutions. The conceptual rationale underlying this work is rooted in the Hopf-Cole transform methods \cite{hopf,cole}, which map solutions of the linear heat equation to those of nonlinear PDEs through cancellation mechanisms (also known as differential cosntraints). These methods have proven powerful and versatile, for instance, in rigorously deriving front position asymptotics under various nonlinearities \( F(u) \) (see eqs. (1.26–1.29) in \cite{ryzhik}). Complementary ideas have emerged from symmetry-based solution-to-solution mappings, with foundational work by \cite{bluman1,bluman2} and further developments in \cite{bluman1,cheviakov}. These broader ideas of solution mappings has precedent in the independent works of Montroll \cite{m1}, with related developments and applications explored in \cite{canosa2,kenkre1,kenkre2}.\\

\noindent While traveling wave ansätze provide a commonly used\cite{ablowitz} and efficient strategy to address PDEs such as \eqref{rde}, typically in the form \( u(x,t) = \Phi(\psi(x - ct)) \), where one derives additional ODE constraints on \( \Phi \) \cite{cherniha,c1,hood2}, our work takes a different route. We do not assume such travelling wave similarity variables. Instead, we adopt a phenomenological perspective by embedding both growth and diffusion within a nonlinear transformation framework akin in spirit to the Hopf-Cole method \cite{hopf,cole}.\\

\noindent Within this framework, the transformation functions play a central role. Specifically, we show that solutions of nonlinear PDEs can be generated by applying nonlinear growth ODEs to solutions of the heat equation, effectively mapping linear dynamics into nonlinear behavior. This defines the core novelty of our approach: transforming heat equation solutions using nonlinear growth dynamics.

\subsection{Objectives}

\noindent The objectives of this work is as follows,
\begin{enumerate}
\item Derive analytical expressions for space-time dependent cell density variable $u(x,t)$ from transformations of heat equation solutions using the growth ODE of Richards model.
\item Apply the analytical solutions $u(x,t)$ using two-variable Hermite polynomials to fit experimental cell density data from a scratch assay, estimating parameters \( m \) and \( b \) in the Richards Growth term.
\item Evaluate the robustness and accuracy of the proposed solutions through parameter estimation, comparing the performance of different ansatzes through stochastic analysis.
\end{enumerate}
The paper is organized as follows : In Model and Methods section we describe in details the methods required for all the 3 objectives as described above. In Results section we describe the parameter estimated from the experimental data. The work is summarized in Conclusion section . Finally some avenues for further research are provided in the Discussion section.

\section{Methods}

{\noindent}We describe the technique which we have used in this section. This technique involves nonlinear transform of the heat equation solutions to construct cell-density field starting from the diffusion equation. Then, cell proliferation model is described which can be connected with the actual experimental observations and the proposed mathematical technique can be applied. Then we described few experimental observations on cell population growth and diffusion followed by parameter estimation of the data retrieved for application.

\subsection{Nonlinear transforms of Heat equation solutions }

\noindent To construct our cell-density field $u(x,t)$ from transformations of the heat equation solutions, we start by defining $h(x,t)$ as the solution of the diffusion operator $T_1 := \partial_t -\partial_{xx} $  that is $T_1h(x,t) = 0$ where the space variable is (standard) scaled by Diffusion constant. Then we define the transformation relations between heat equation solution $h(x,t)$ and the cell-density variable $u(x,t)$ as:
\begin{equation}
G(u(x,t)) = g(f(t)h(x,t))
\label{Geqn}
\end{equation}
where the two transformation functions $G$ and $g$ are defined to have well behaved inverses and $f(t)$ is an arbitrary function yet to be determined. We apply $T_1 := \partial_t -\partial_{xx} $ to \eqref{Geqn} which gives us a PDE $$ \Omega_1 (G,G',G'', g,g',g'',u_x,u_{xx}, u_t,h_x,h_{xx},h_t)=0 $$ and in the next step we make use of the nonlinear growth ODE \eqref{grow} to impose conditions on  $G, G',G''$ and $g,g',g''$ etc.  for reduction(cancellation) of terms in $\Omega_1$ as done in Hopf-Cole methods (pp.230-231 in \cite{cole}). For the detailed derivations we refer to the supplementary material.\\

\noindent The primary aim is to derive two distinct relations between $u(x,t)$ and $h(x,t)$ by imposing conditions on the two arbritrary functions $g$ and $G$. These are achieved through the two ansatze as presented below:\\

\noindent \textbf{Ansatz I:} This ansatz is given by the non-linear growth ODE on the function $g$ in the form  
\begin{equation}
\frac{dg}{dh}=F(g) 
\label{ans1}
\end{equation} together with $G(u) = u$ as identity mapping thus we obtain its inverse function $G^{-1}(u)=u$ and set without any loss of generality $f(t) = 1$ to get 
\begin{eqnarray}
u = g(h)
\end{eqnarray}
This completes our Ansatz I. By using the relation $u = g(h)$ and the ODE \eqref{ans1} we get $u$ in terms of the heat equation solution $h$ as :
\begin{equation}
\int \frac{du}{F(u)} = h 
\label{res1}
\end{equation}
where, $h(x,t)$ satisfies $h_t= h_{xx}$.  On performing the integration and inverting the expression for $u$ we get an expression in term of $h$. A crucial assumption here is that the inverse function of $g$ exists and can be written in closed form. \\

\noindent \textbf{Ansatz II} :In this Ansatz the condition on G is given by 
\begin{equation}
    \frac{d}{du}\big[\log_e G(u)\big]  = \frac{1}{F(u)}
    \label{ans2}
\end{equation} and together with $g(fh)=fh$ as identity mapping which gives $ g^{-1}(fh)= fh $ and we set $f(t)=e^t$. This reduces the central ansatz \eqref{Geqn} as $G(u) = fh$ .
This completes our Ansatz II.\\
\noindent We get for an analytical expression of $u(x,t)$ the form as a solution of an integral equation:
\begin{eqnarray}
\exp \left[\int \frac{du}{F(u)}  \right]=fh
\label{res2}
\end{eqnarray}
where, $h(x,t)$ satisfies $h_t= h_{xx}$ .
\noindent Thus as before again on performing the integration and inverting the expression for $u(x,t)$ we get an expression in term of $h(x,t)$.\\

\noindent Thus these ansatze yielded two key relations connecting the cell-density function $u(x,t)$ with a solution to the heat equation $h(x,t)$, :

\begin{itemize}
    \item From \textbf{Ansatz I:} \quad \( \displaystyle \int \frac{du}{F(u)} = h \)
    \item From \textbf{Ansatz II:} \quad \( \displaystyle \exp\left[\int \frac{du}{F(u)}\right] = fh \)
\end{itemize}
\noindent These relations provide two distinct ways for modeling $u(x,t)$ based on solutions of the  heat equation $h(x,t)$.
\begin{table*}[htbp]
    \caption{Summary of variables and parameters used in the modeling and analysis.}
    \centering
    \begin{tabular}{l l p{10cm}}
        \toprule
        \textbf{Symbol} & \textbf{Name} & \textbf{Description} \\
        \midrule
        \hline
        \multicolumn{3}{l}{\textit{Model Variables}} \\
        \hline\\
        \midrule
        \( u(x,t) \) & Cell density & Cell population density, \( u \in [0,1] \), as a function of space \( x \in \mathbb{R} \) and time \( t \in [0, \infty) \)\\
        \( x \) & Spatial coordinate & Position in one-dimensional space, \( x \in \mathbb{R} \), defining the spatial domain for cell migration. \\
        \( t \) & Time & Time variable, \( t \in [0, \infty) \) for tracking the temporal evolution of cell density. \\
        \midrule
        \hline
        \multicolumn{3}{l}{\textit{Growth Parameters (Richards Model)}} \\
        \hline\\
        \midrule
        \( a \) & Growth rate & Intrinsic growth rate of the cell population, \( a > 0 \), scaled with the time variable controlling the speed of cell proliferation in the reaction term \( F(u) \). \\
        \( b \) & Maximum relative density & Parameter related to the maximum dimensionless relative density $u(x,t)$ in the Richards growth term. \( b > 0 \)  \\
        \( m \) & Nonlinearity exponent & Exponent controlling the nonlinearity of the growth term, \( m > 0 \) adjusting different dynamics to the growth curve from logistic (\( m = 1 \))  \\
        \midrule
        \hline
        \multicolumn{3}{l}{\textit{Initial Condition Parameters (Scratch Assay)}} \\
        \hline\\
        \midrule
        \( \eta \) & Unscratched region cell density & Parameter controlling the maximum cell density in the initial condition \( u(x,0) \) defining the background density outside the scratched region. \\
        \( \alpha \) & Depth parameter & Parameter controlling the depth of the cleared region in the initial condition, \( \alpha \in (0,1) \) determining the reduction in cell density at the center of the scratch. \\
        \( \beta \) & Width parameter & Parameter controlling the width of the transition zone in the initial condition, \( \beta > 0 \) governing the spatial extent of the cleared region. \\
        \midrule
        \hline
        \multicolumn{3}{l}{\textit{Optimization and Experimental Parameters}} \\
        \hline\\
        \midrule
        \( L \) & Spatial domain half-length & Half-length of the spatial domain defining the solution domain \( x \in \Omega := [-L, L] \) for modeling. \\
        \( t_{\text{max}} \) & Maximum time & Maximum time for evaluation of analytical solutions i.e. \( t \in \Omega := [0, t_{max}] \) . \\
        \bottomrule
    \end{tabular}
    \label{table1}
\end{table*}

\subsection{Cell proliferation model and diffusion }

\noindent The total cell density at a given point of space $(x)$ at any given time $(t)$ is modeled by the help of the variable $u(x,t)$ constructed from a non-linear transformation of the heat equation solution. $u(x,t)$ satisfies the solutions (\eqref{res1} and \eqref{res2})provided from the ansatz 1 and 2, the full solutions are given in Table \ref{table2}.\\

\noindent The two subpopulations representing the cells in the different stage of the cell cycles are colored differently as shown in \cite{vittadello} i.e. red for G1 phase and green for S/G2/M phases respectively. These subpopulations are also constructed using the expressions from (p.2 in \cite{baker2}). They are :
\begin{align}
\rho_1(x,t) &= \frac{\Gamma_2}{\Gamma_1 + \Gamma_2} \, u(x,t), \\
\rho_2(x,t) &= \frac{\Gamma_1}{\Gamma_1 + \Gamma_2} \, u(x,t)
\end{align}
where $\rho_1$ represents the cells in G1 phase(colored red) and $\rho_2$ represents the cells in S/G2/M phase(colored green).

\subsection{Data : Scratch Assay experiment}
\noindent The experimental setup of the scratch assay as described in \cite{vittadello}. The non-dimensional cell density $u(x,t)$ is obtained by scaling the number of cells with the area of the column(the observed area is divided into columns of equal width) and the theoretical maximum packing of cells given by $K=0.004$ $\mu m^{-2}$ as is given in \cite{vittadello,baker1}. The data tables \cite{data} from the scratch assays are given at four time points after the scratch is made: 0, 16, 32, and 48 h for the 1205Lu cell line, see (scatter black dots) Fig. \ref{fig_lu_error}, (e-f); and for the WM983C cell line see Fig. \ref{fig_wm_error} (e-f). For the C8161 cell line the data was given for 0, 6, 12, and 18 h see Fig. \ref{fig_c8_error}(e-f). It can be distinctly observed that the cell proliferates as well as they diffuse into the gap as time progresses. \\

\noindent We consider a scratch assay, where a region of a cell-layer is cleared, and cells repopulate the gap over time. The initial condition is modeled as a smooth function, as shown in the images by :
\begin{equation}
    u(x,0) = \frac{1}{\eta + \frac{\alpha}{1 - \alpha} e^{-\frac{x^2}{2 \beta^2}}},
    \label{init}
\end{equation}
where \( \eta > 0 \) controls the maximum density otsude the scratched area, \( \alpha \in (0,1) \) governs the depth of the cleared region, and \( \beta > 0 \) determines the width of the transition. For detailed derivation please refer to the supplementary material.\\

\noindent This initial condition mimics the experimental setup, with a low-density region centered at \( x = 0 \). For the analysis we have estimated and fixed from data the initial conditions for the various cell lines as follows: for Ansatz 1 we have for 12505Lu cell line $\alpha= 0.998, \beta=6.2 , \eta =10 $ WM983C cell line $\alpha= 0.998, \beta=8 , \eta = 10$ and for C8161 cell line $\alpha= 0.9905, \beta=6.2 , \eta =10 $ . For Ansatz 2 we have for 12505Lu cell line $\alpha=0.99 , \beta=5 , \eta =2 $ WM983C cell line $\alpha=0.989 , \beta=8 , \eta = 2$ and for C8161 cell line $\alpha= 0.99, \beta=5 , \eta =2 $ . We can see that the initial conditions for cell line WM983C the width of the scratched region is much wider than the other two. These values are obtained from fitting the function \eqref{init} to the $t=0$ data provided in the sheets\cite{data}.

\subsection{Solutions using Hermite polynomials}

\noindent The initial conditions taken in the above section cannot be used in the heat kernel to obtain closed form solutions hence we take the functional approximation route by representing solutions through orthogonal polynomials: the 2 variable Hermite polynomials ${}^{(2)}H_n(x, t)$. For full details of the orthogonality and other properties we refer to the work in (pp: 74-78 in \cite{dattoli}).\\

\noindent The solution of heat equation can be written in terms of the multi-variable Hermite polynomials ${}^{(2)}H_n(x, 2t)$ and the Maclaurin series of the initial condition $h(x,0)=\phi(x)$ :
\begin{eqnarray}
h(x,t)= \sum_{n}^{}\frac{\phi^{(n)}(0)}{n!} \left[{}^{(2)}H_n(x, 2t)\right]
\end{eqnarray}
where,  $\phi^{(n)}(0)$ denotes the $n^{th}$ derivative evaluated at $x=0$ i.e the Maclaurin series of $\phi$. \\

\noindent For our work we have truncated the series at $m=4$ as the higher order terms are $O(\frac{1}{n!})$. Hence we have used the following approximate solutions :
\begin{eqnarray}
h(x,t)= \sum_{n=0}^{m}\left( \Theta_n \right) \left[^{(2)}H_n(x,2t)\right]
\end{eqnarray}
where,  $\Theta_n = \frac{\phi^{(n)}(0)}{n!}$

\subsection{Parameter Estimation}

\begin{figure*}[htbp]
    \centering
\includegraphics[scale=0.45]{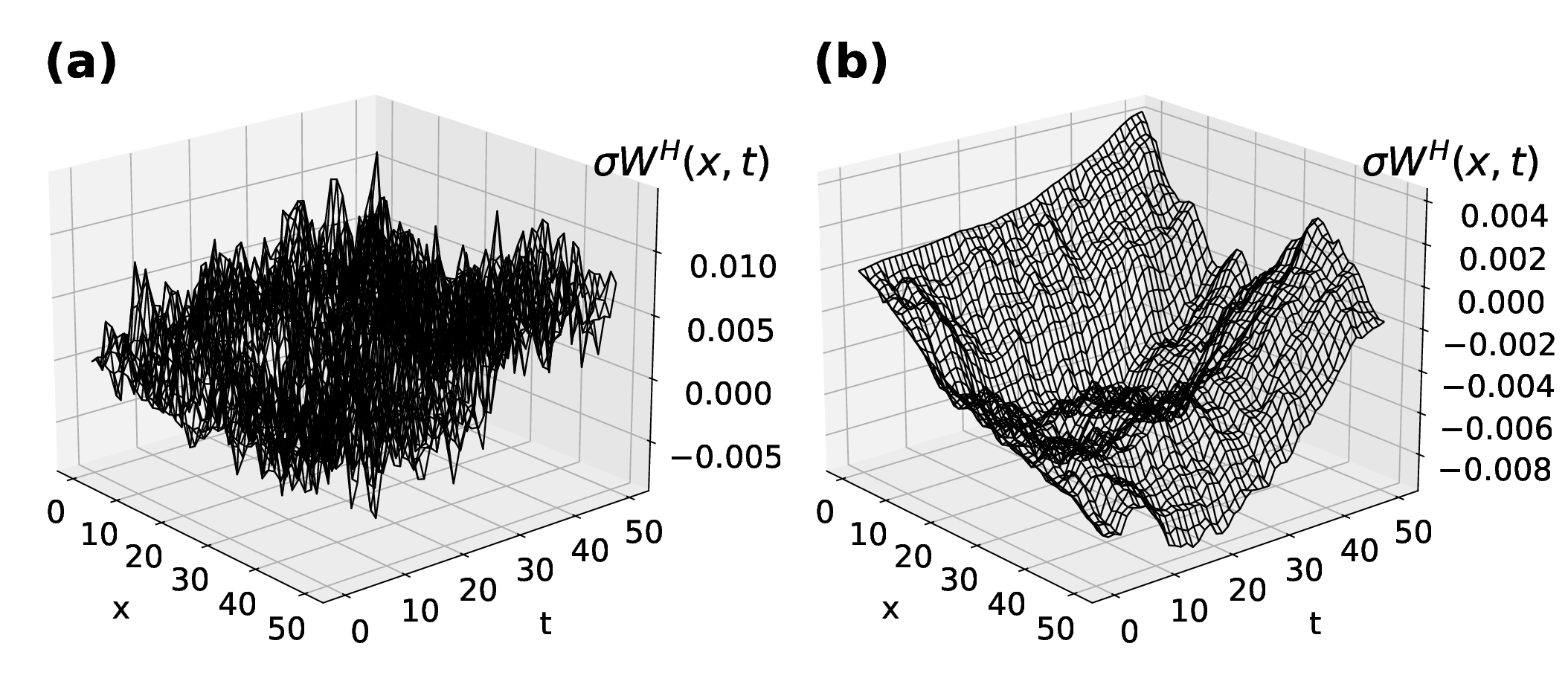}
    \caption{ A single realization of $\sigma W^H(x,t)$:= the fBM fields with different Hurst exponents (H). (a)H=0.3, (b)H=0.8 For all the of the figures the $\sigma = 0.006$}
    \label{figfbm}
\end{figure*}  
\noindent We do the parameter estimation in two steps. In first step we assume the residual error to be given by the expression:
\begin{equation}
   u^{obs}(x,t) = u^{ana}(x,t;\mathbf{\Theta})+ \epsilon(x,t)
   \label{err1}
\end{equation}
where, $\epsilon$ are the residuals(error) to be calculated from the given model and the data and parameter set is defined as $\mathbf{\Theta}:=\{m,b\}$. As a first step we minimize the norm of this residuals to get the optimal parameter values. A Gaussian distribution for the residuals is assumed. This forms the basis of the log-likelihood function taken. For detailed expression we refer to the work in (pp. 1774) in \cite{baker3}. \\

\noindent For a more robust analysis we take a new method, a more general approach to find out the covariance structure present in the data by assuming that the residuals can be modeled by a $W^{H}(x,t)$ is a fractional Brownian motion field whose space-time covariance given by the Hurst exponent (H) and strength controlled by sigma ($\sigma$). For details of this type of noise we refer to \cite{tindel}. We have shown the fBM field for two values of H in the figure \ref{figfbm}.\\
 
\noindent Thus we modify \eqref{err1} and we can write the equation for residuals as : 
\begin{equation}
    u^{obs}(x,t) = u^{ana}(x,t;\mathbf{\Theta})+ \sigma W^{H}(x,t) \label{err2}
\end{equation} 
where, $u^{obs}(x,t)$ is the data, $u^{ana}(x,t) $ is the solution of the PDE model generated by the $\mathbf{\Theta}$ parameter set and and $W^{H}$ is a random field  with Hurst exponent(H) (p. 193 in \cite{tindel}). Our goal is the estimation of the $\sigma$, H value and the $\mathbf{\Theta}$ parameter set that best explains the data. i.e. minimizes residuals as defined below.\\

\noindent Given $[{x_j; j=1, 2...M}] \times [ {t_i; i= 1...N} ]$ we find the total residual error by the following steps:\\
\begin{enumerate}
\item We first compute for a fix parameter set $\mathbf{\Theta}$ the norm across each space point(i.e., per time $t_i$):
\begin{eqnarray}
\epsilon^{t_i} = \sqrt{ \sum_{x_j=1}^{M} (\epsilon_{x_j}^{t_i})^2 }
\end{eqnarray}
where, 
\begin{eqnarray}
(\epsilon_{x_j}^{t_i}) : = (u^{ana}(x_j,t_i;\mathbf{\Theta}) + \sigma W^{H}(x_j,t_i)) - u^{obs}(x_j,t_i)  
\end{eqnarray}
\item Then we find the total residual error as the norm across all $t_i$:
\begin{equation}
    \text{Error} = \sqrt{ \sum_{i=1}^{N} (\epsilon^{t_i})^2 } 
\end{equation} 
This total residual Error is minimized to get the optimal parameter set $\mathbf{\Theta}, H ,\sigma$.
\end{enumerate}

\begin{figure*}[htbp]
    \centering
    \includegraphics[width=\linewidth]{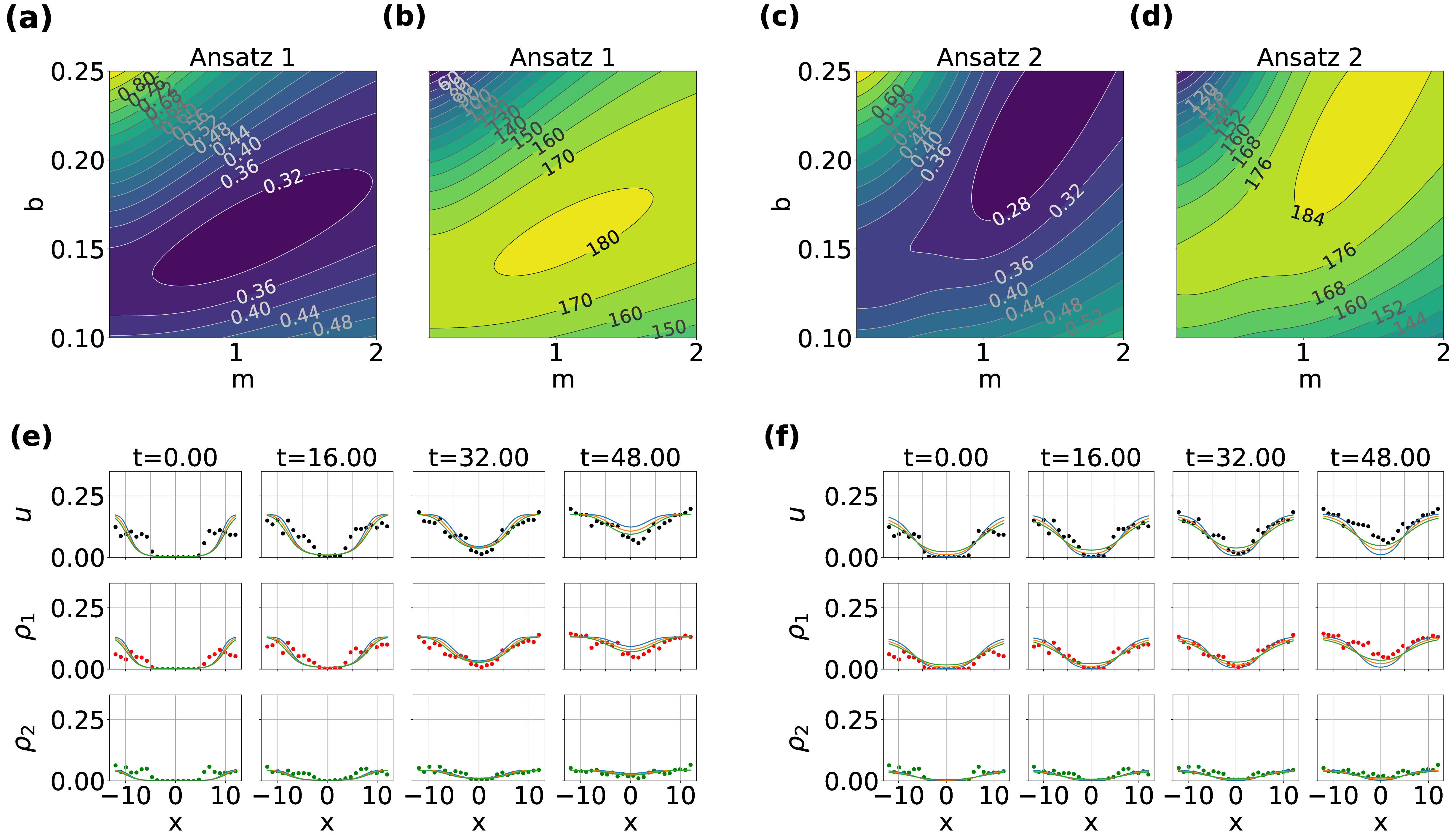}
    \caption{Residual Error, log-likelihood surfaces and estimation of parameters for the 12505Lu cell line with the Richards growth model. Contour plots: (a, c)  the norm of the total residuals(error) surface and (b,d) log-likelihood surface as a function of Richards parameters \( b \)  and \( m \) for Ansatz I and Ansatz II, respectively. (e-f)Experimental data (dots) and fitted analytical solutions (lines) at time points \( t = 0.0, 16.0, 32.0, 48.0 \) hours (columns) for Ansatz I and Ansatz II, respectively, showing total cell density \( u \) (top row), G1-phase(red colored cells) cell density \( \rho_1 \) (middle row), and S/G2/M-phase(green colored cells) cell density \( \rho_2 \) (bottom row).}
    \label{fig_lu_error}
\end{figure*}

\begin{figure*}[htbp]
    \centering
    \includegraphics[width=\linewidth]{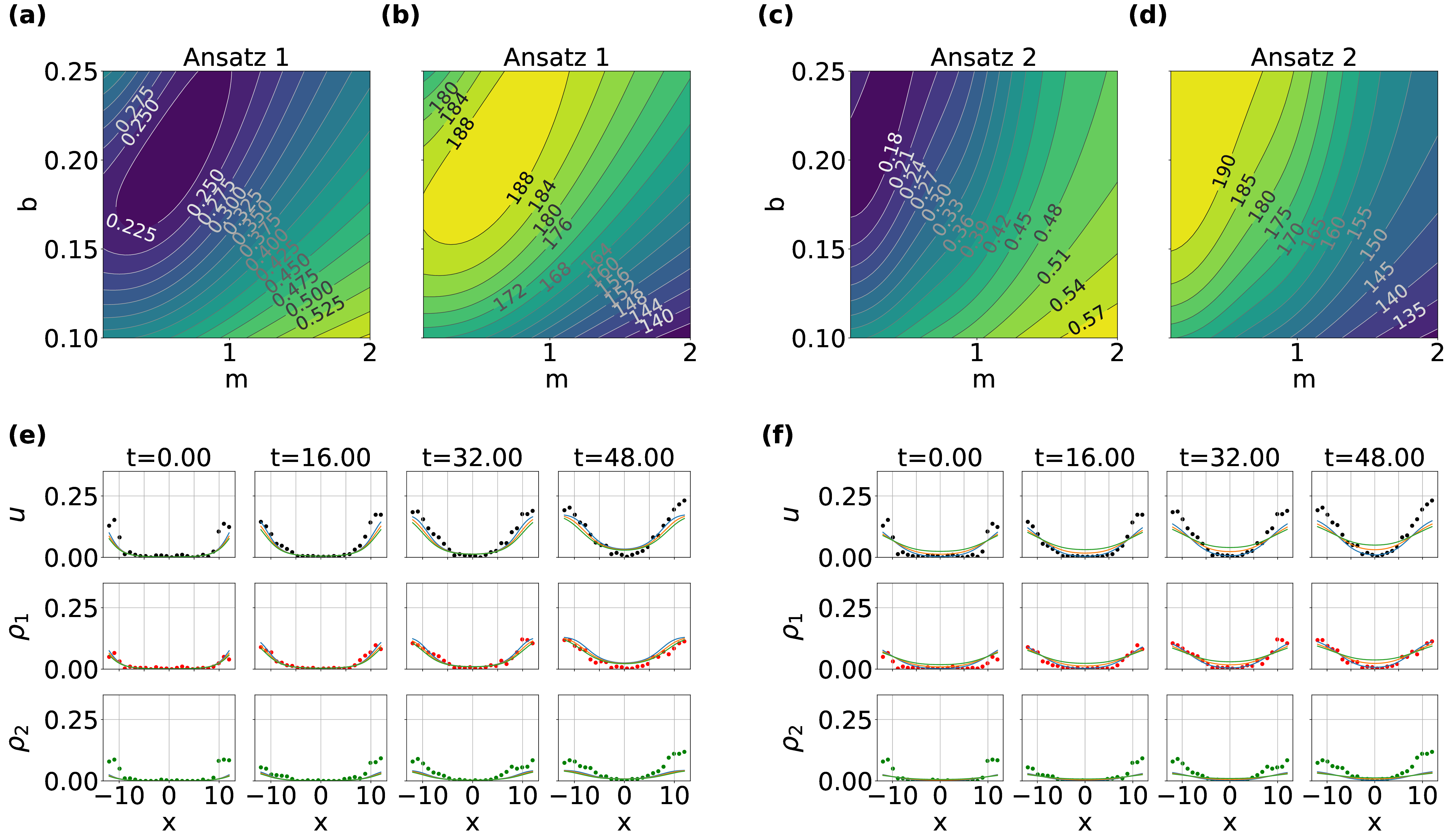}
    \caption{Residual Error, log-likelihood surfaces and estimation of parameters for the WM983C cell line with the Richards growth model. Contour plots: (a, c)  the norm of the total residuals(error) surface and (b,d) log-likelihood surface as a function of Richards parameters \( b \)  and \( m \) for Ansatz I and Ansatz II, respectively. (e-f)Experimental data (dots) and fitted analytical solutions (lines) at time points \( t = 0.0, 16.0, 32.0, 48.0 \) hours (columns) for Ansatz I and Ansatz II, respectively, showing total cell density \( u \) (top row), G1-phase(red colored cells) cell density \( \rho_1 \) (middle row), and S/G2/M-phase(green colored cells) cell density \( \rho_2 \) (bottom row).}
\label{fig_wm_error}
\end{figure*}

\begin{figure*}[htbp]
    \centering
    \includegraphics[width=\linewidth]{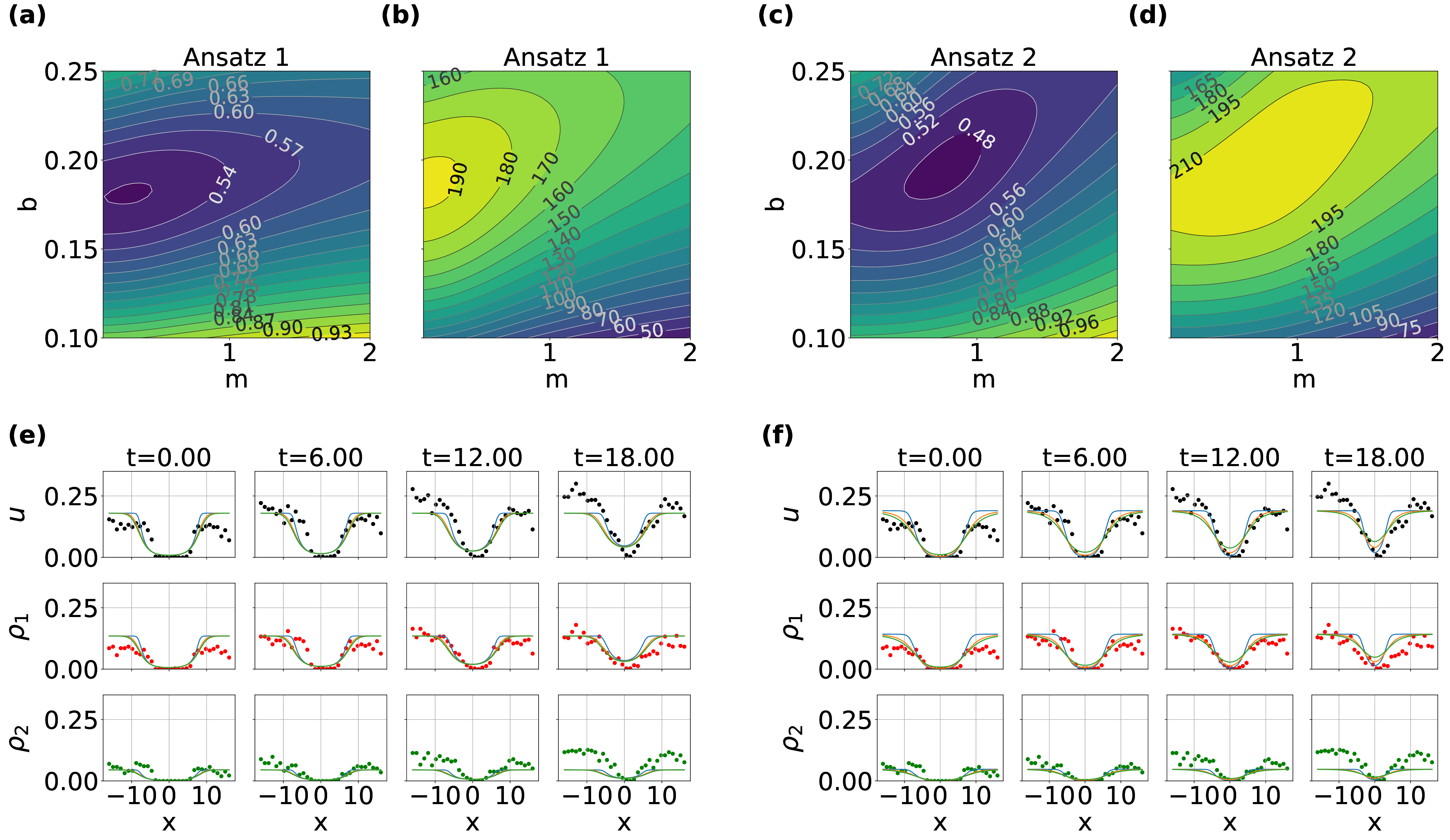}
    \caption{Residual Error, log-likelihood surfaces and estimation of parameters for the C8161 cell line with the Richards growth model. Contour plots: (a, c)  the norm of the total residuals(error) surface and (b,d) log-likelihood surface as a function of Richards parameters \( b \)  and \( m \) for Ansatz I and Ansatz II, respectively. (e-f)Experimental data (dots) and fitted analytical solutions (lines) at time points \( t = 0.0, 6.0, 12.0, 18.0 \) hours (columns) for Ansatz I and Ansatz II, respectively, showing total cell density \( u \) (top row), G1-phase(red colored cells) cell density \( \rho_1 \) (middle row), and S/G2/M-phase(green colored cells) cell density \( \rho_2 \) (bottom row).} 
    \label{fig_c8_error}
\end{figure*}

\section{Results}\label{results}
\begin{table*}[htbp]
		\caption{Summary table of solutions(Richards Model).}
		\centering 
		\begin{tabular}{ p{5cm} p{5cm}}
			\hline 			
			     \quad \textbf{Ansatz I} & \textbf{Ansatz II}
             \\	
			\hline		
            \\
		   $u(x,t)= \frac{1}{\left[b^{-\frac{1}{m}} +  e^{-\frac{ah(x,t)}{m}}\right]^{m}}$ & $u(x,t)= \frac{1}{\left[b^{-\frac{1}{m}} +\left[e^t h(x,t)\right]^{-\frac{a}{m}}\right]^{m}}$  \\
            \\
            \hline
			\hline
			
		\end{tabular}
        \label{table2}
\end{table*}
{\noindent}We, now, present our results. As summarized in the Table \ref{table2} we provide the solutions from \eqref{res1} and \eqref{res2} for the Richards function\cite{nelder,richards} $$F(u) = au\bigg( 1-\bigg( \frac{u}{b}\bigg)^{\frac{1}{m}}\bigg)$$ In the table the function of integration is omitted without loss of generality. We use these solutions to estimate the parameters as described below. Additionally we keep $a =1$ fixed for all the subsequent analysis.

\subsection{Estimates of parameters}
\noindent The parameters to be estimated from the data are: $\mathbf{\Theta}= \{m,b\}$ the Richards exponent and the maximum relative cell-density. We took two approaches for this:
\begin{enumerate}
\item Estimation of the parameters $\mathbf{\Theta}= \{m,b\}$ using Euclidean norm of the residuals(error) and using the maximum of the log-likelihood surface as described by assuming the form \eqref{err1}.
\item Estimation of the same parameters $\mathbf{\Theta}= \{m,b\}, H, \sigma$ where the parameters $H$ is the Hurst exponent  and noise strength $\sigma$ by addition of a fBM field to the analytical solution and subsequent computation of the error norm as described by assuming the form \eqref{err2}.
\end{enumerate}

\subsubsection{Estimation using residual error and log-likelihood surfaces}

\noindent First we take up \eqref{err1}, where we assume a Gaussian distribution of the residual Errors. We plot the error surface by varying the desired parameters m and b. The contour plots of error surfaces for different cell lines for both the ansatze are provided in the figures: 12505Lu cell line in Fig. \ref{fig_lu_error}(a,c), WM983C cell line in Fig. \ref{fig_wm_error}(a,c) and for C8161 cell line in Fig.\ref{fig_c8_error}(a,c). For all the above cell lines and for both the ansatze the error surface has a distinct minimum band as clearly seen from the contours plots. The fitting seems to be working robustly in the cases of 12505Lu and WM983C cell lines while there is little deviations in the case of C8161.\\

\noindent To confirm the above regions we also show the log-likelihood surfaces for 12505Lu cell line in Fig. \ref{fig_lu_error}(b,d), WM983C cell line in Fig. \ref{fig_wm_error}(b,d) and for C8161 cell line in Fig.\ref{fig_c8_error}(b,d). Clearly the maximum of the log-likeliood surfaces coincides with the residual error surfaces as seen in the figures. The Richards growth model, with \( m \) values near 1, effectively captures the logistic-like growth behavior across different cell lines. Let us also remark that the values we have estimated match the growth observed for cellular processes given in (p.11 Fig.7 of \cite{maini}) through completely different methods.

\subsubsection{Estimates of parameters from stochastic analysis }
\noindent For the stochastic case a fractional Brownian motion (fBm) field $W^{H}(x,t)$  is used via the error formula \eqref{err2}: 
 \begin{equation}
    u^{obs}(x,t) = u^{ana}(x,t;\mathbf{\Theta})+ \sigma W^{H}(x,t) 
\end{equation} 
where \( \sigma \) is the noise strength and \( H \) is the Hurst exponent, allowing us to estimate \( b \), \( m \), \( H \), and \( \sigma \) while capturing stochastic variability in cell dynamics. The H values for the search region is kept as $H \in [0.3,0.8]$ in line with the existence results in \cite{tindel}.
\begin{figure*}[htbp]
    \centering
    \includegraphics[width=\linewidth]{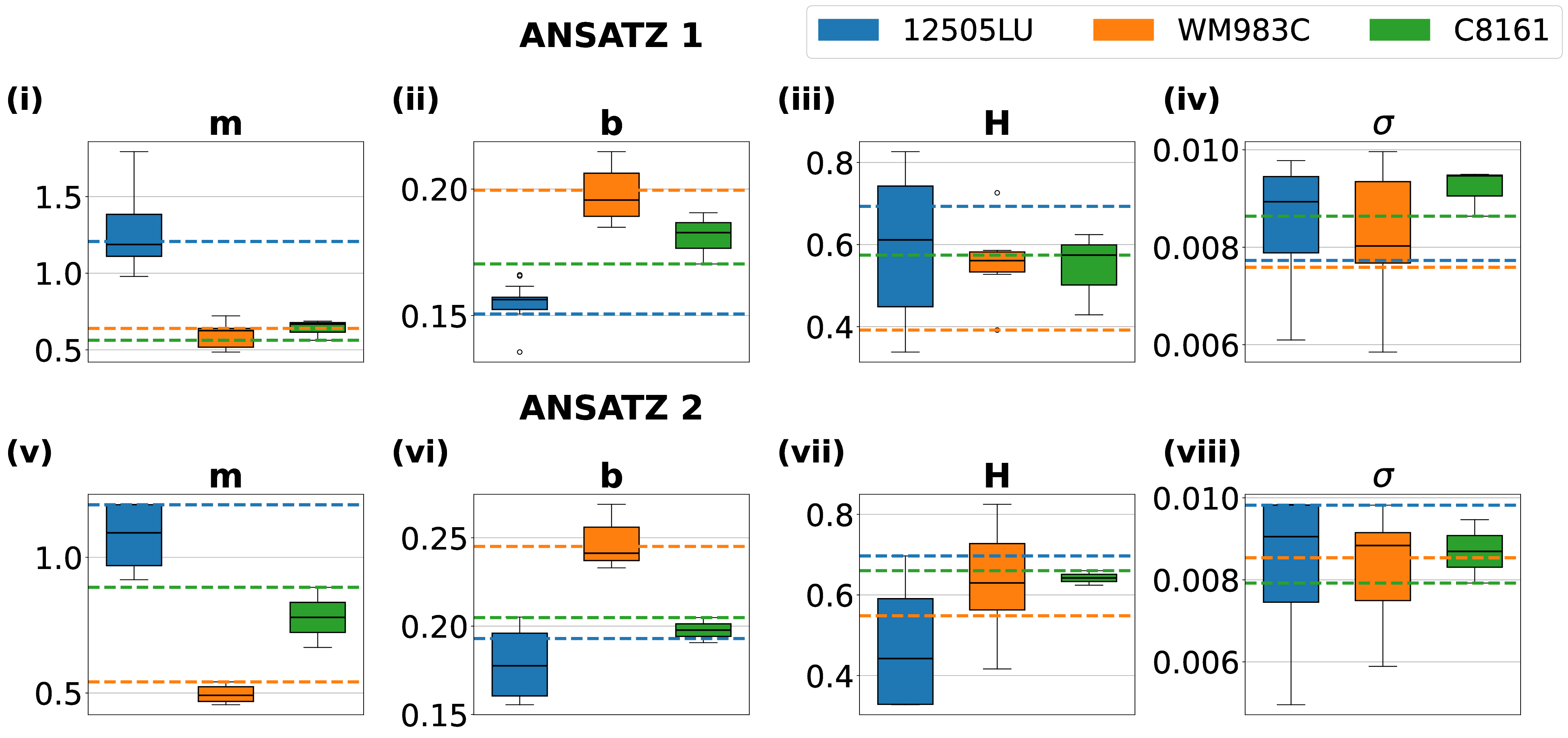}
    \caption{Parameters estimated from the stochastic analysis using fBM field. Boxplots of model parameters ($m,b, H, \sigma$) that fall within a narrow window of the global minimum residual error (i.e. the subset formed by [minimum error, minimum error + 0.005] ) for two ansatze top row(i-iv) and Ansatz 2 bottom row (v-viii) across three cell lines. The dash line indicates the parameter values at exactly the global minimum of total residual error. The search grid for the parameters used 70,000 parameter tuples randomly chosen (uniformly distributed) between biologically relevant ranges.}
    \label{parameters}
\end{figure*}

\noindent The parameters estimated and their distributions are plotted in Figure \ref{parameters}. To analyze the behavior of model parameters corresponding to minimal residual error, we extracted the subset of parameter estimates from three different datasets (12505Lu, WM983C, and C8161) that fall within a narrow window above the global minimum (i.e. the subset formed by [minimum error, minimum error + 0.005] ). For each parameter ($m$, $b$, $H$, and $\sigma$), we plotted comparative box-plots grouped by datasets i.e. cell lines. These are organized into two rows: the first row for Ansatz 1 and the second for Ansatz 2 based subsets. Additionally, horizontal dashed lines are overlaid to indicate the exact parameter values corresponding to the absolute minimum error within each dataset. This visualization highlights the variability of parameter distributions near optimal solutions and helps contrast the behavior across different data regimes.

\begin{figure*}[htbp]
    \centering
    \includegraphics[width=\linewidth]{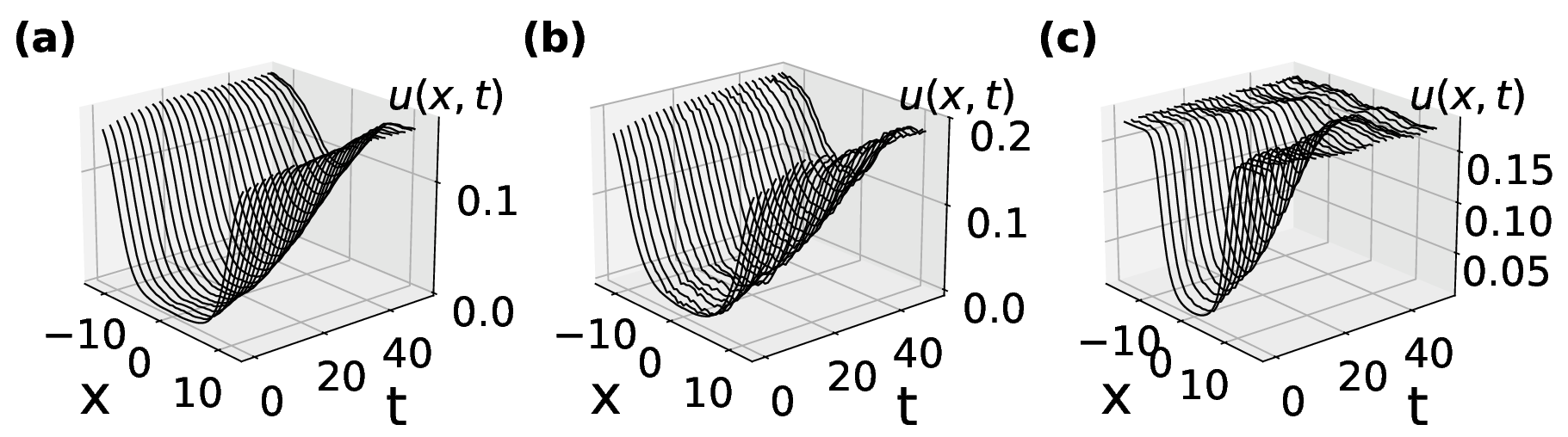}
    \caption{Surfaces from optimal parameters for Ansatz 1. Parameters are described in main text \eqref{param1}.}
    \label{sa1}
\end{figure*}

\begin{figure*}[htbp]
    \centering
    \includegraphics[width=\linewidth]{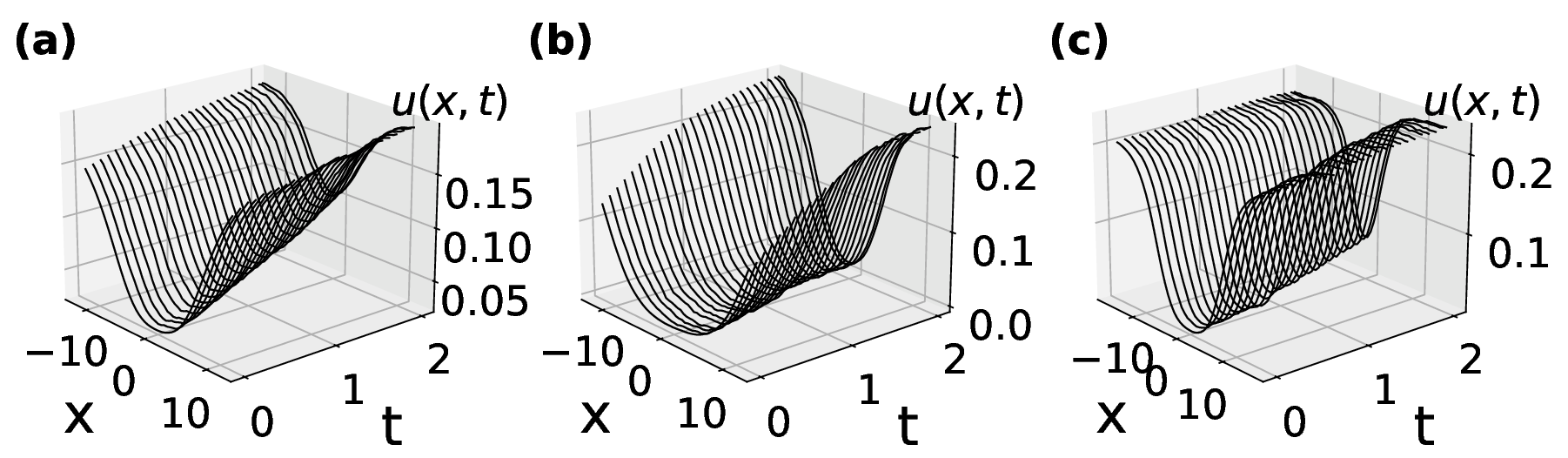}
    \caption{Surfaces from optimal parameters for Ansatz 2. Parameters are described in main text \eqref{param2}.}
    \label{sa2}
\end{figure*}

\noindent We plot the evolved surfaces in Figures \ref{sa1} and \ref{sa2} using the parameters($b,m,H, \sigma$) inferred above (dashed lines in figure \ref{parameters}) for Ansatz 1 and 2 respectively. As can be seen, the surfaces evolve distinctly depending on the initial conditions and parameters sets. For Ansatz 1 the surface is evolved for $t = [0,50]$. The parameters used for the various cell lines using Ansatz 1 are :
\begin{align}
\text{12505LU:} &\quad m = 1.207,\ a = 1,\ \eta = 10,\ \alpha = 0.998,\ b = 0.15,\ \beta = 6.2,\ L = 12,\ H = 0.69,\ \sigma = 0.007 \nonumber \\ 
\text{WM983C:} &\quad m = 0.639,\ a = 1,\ \eta = 10,\ \alpha = 0.989,\ b = 0.19,\ \beta = 8.0,\ L = 12,\ H = 0.39,\ \sigma = 0.007 \nonumber \\ 
\text{C8161:} &\quad m = 0.5,\ a = 1,\ \eta = 10,\ \alpha = 0.9905,\ b = 0.17,\ \beta = 6.2,\ L = 16,\ H = 0.574,\ \sigma = 0.008
\label{param1}
\end{align}

\noindent For Ansatz 2 the surface is evolved for $t = [0,2]$. The parameters used for the various cell lines he parameters used for the various cell lines Ansatz 2 are:
\begin{align}
\text{12505LU:} &\quad m = 1.19,\ a = 1,\ \eta = 2,\ \alpha = 0.99,\ b = 0.19,\ \beta = 5,\ L = 12,\ H = 0.69,\ \sigma = 0.009 \nonumber\\
\text{WM983C:} &\quad m = 0.54,\ a = 1,\ \eta = 2,\ \alpha = 0.989,\ b = 0.24,\ \beta = 8,\ L = 12,\ H = 0.548,\ \sigma = 0.008\nonumber \\
\text{C8161:} &\quad m = 0.75,\ a = 1,\ \eta = 2,\ \alpha = 0.99,\ b = 0.21,\ \beta = 5,\ L = 16,\ H = 0.626,\ \sigma = 0.009
\label{param2}
\end{align}
The time to which the surfaces are evolved are different as the nature of the ansatze are very distinct. Ansatz 1 does not involve a modulating function $f(t)$ while Ansatz 2 has $f(t) = e^t$ . Thus the time scales in both the ansatz are distinct. This distinction can also be marked in the estimates of $m, b$ parameters in Figure \ref{parameters}(i-ii,v-vi).

\section{Discussion}\label{discussion}

\noindent In this work, through careful parameter estimation, we identified optimal ranges for the nonlinearity exponent \( m \) and \( b \) that minimize errors, achieving a close match between analytical predictions and experimental measurements. Our findings validate the efficacy of the Richards growth model to explain the growth of the cells line given in the data. The analytical solutions presented here offer a computationally efficient and theoretically grounded approach to modeling cell growth and diffusion, with broad implications for understanding biological systems. \\

\noindent Despite these strengths, our study has two major limitations that warrant consideration. Firstly, the analytical solutions rely on the assumption that \( h(x,t) \) can be accurately represented by a finite Hermite polynomial expansion (up to fourth order in this work). We observe that on taking higher orders there is no significant change in the solutions as the terms contribute very less to the total sum i.e. they go to zero by the order O($\frac{1}{n!}$). Second limitation of this work is that the initial conditions are kept fixed. Optimizing by allowing variability in iniital condition functions require more research and is still open avenue. Further if we can find any other smooth function( resembling the initial condition as given in the data) which can be convoluted with the heat kernel we can obtain the closed form solution and it would yield more precise comparison. Future work may also include two statistical comparative analysis: firstly, bootstrap methods for comparing how usage of small sample data effects the estimates and secondly comparing model information criteria using Akaike Information Criterion (AIC) etc. as done in works of \cite{baker3} to find out how these models generate perform and provide a metric in the model space to compare models.

\section{Conclusion}\label{conclusion}
\noindent This study presents a novel analytical framework for modeling cell growth using the the Richards growth model ODE and diffusion through leveraging nonlinear invertible transformations of heat equations. Thus we give alternative ways to construct density fields to the ones already widely studied in the literature. We show that these solutions, summarized in Table \ref{table2}, provide a robust and computationally efficient means to capture the spatiotemporal dynamics of cell populations.\\

\noindent A key contribution of this work is the validation of these analytical solutions against experimental data from a scratch assay, as reported in \cite{vittadello}. The experimental dataset, comprising cell density measurements for the 12505Lu, WM983C and C8161 cell line at four time points across spatial points, was used to assess the accuracy of our solutions. By fitting the analytical solutions to the data, we estimated the key parameters \( m \) (nonlinearity exponent) and \( b \) (maximum relative density achieved) through a grid search over \( m \) and \( b \) parameter values. As previously remarked we recover for $m$ a narrow band between values [0.5-1.5] which was also estimated for such cellular processes in \cite{maini} albeit from entirely different methods. We also observe that we achieve the same results as mentioned in pp.929-930 \cite{baker1} where it was stated that the cellular proliferation and diffusion are adequately described by the simpler models. Thus a simple model such as Richards explains data well(the one employed in this work) instead of invoking density dependent effects. Probably more data is required for entangling more complicated effects.\\

\noindent The success of our analytical solutions in replicating experimental results has significant implications for mathematical biology, particularly in understanding cancer cell dynamics and tissue regeneration, where scratch assays are widely used to study cell migration and growth \cite{vittadello}. The Richards growth term, which allows for flexible modeling of nonlinear growth behaviors through the parameter \( m \). Our findings suggest that the optimal \( m \) values near 1 align with logistic-like growth, consistent with biological observations of near exponential proliferation in these cell lines.\\\\

\noindent\textbf{Author Contributions:} Conceptualization, P.M., S.K., S.R.S., R.K.B.S.; Analysis and numerical simulations, P.M. , S.K.; Writing—original draft preparation P.M., S.K., R.K.B.S.; Writing—review and editing, P.M., S.K., S.R.S., R.K.B.S.; Supervision, R.K.B.S.; All authors have read and approved the manuscript.\\

\noindent\textbf{Acknowledgment:} P.M. acknowledges CSIR for Senior Research Fellowship(Direct) funding. S.K. acknowledges ICMR for Senior Research Fellowship funding. R.K.B.S. acknowledges JNU and DBT BIC for facilities.\\

\noindent\textbf{Data Availability Statement:}
Data used is open access and available in \url{https://www.cell.com/cms/10.1016/j.bpj.2017.12.041/attachment/1a1897ab-08fe-41fe-92b1-97c4e3c02344/mmc2.xlsx} in the supplemental material of the work \cite{vittadello}. \\

\noindent\textbf{Conflicts of Interest:} The authors declare no conflicts of interest.

\end{document}